\documentclass[acmsmall]{acmart}
\newcommand{\chatgpt}[1]{{\leavevmode\color{black}#1}}%
\usepackage{numprint}%
\nprounddigits{5}%
\npdecimalsign{.}%
\npthousandsep{}%
\npfourdigitnosep%
\usepackage{caption}%
\usepackage{subcaption}%
\setcopyright{rightsretained}%
\copyrightyear{2023}%
\acmYear{2023}%
\acmDOI{10.48550/arXiv.2303.13530}%
\acmConference[CHI ’23 Workshop on Generative AI and HCI]{GenAI Workshop 2023}{Apr 28, 2023}{Virtual}
\acmISBN{}

\begin{document}

\title{\chatgpt{%
Text-to-Image Generation: Perceptions and Realities}}%

\author{Jonas Oppenlaender}
\email{jonas.x1.oppenlander@jyu.fi}
\affiliation{%
  \institution{University of Jyv\"askyl\"a}
  \city{Jyv\"askyl\"a}
  \country{Finland}
  \postcode{40014}
}


\author{Aku Visuri}
\email{aku.visuri@oulu.fi}
\affiliation{%
  \institution{University of Oulu}
  \city{Oulu}
  \country{Finland}
  \postcode{90570}
}

\author{Ville Paananen}
\email{ville.paananen@oulu.fi}
\affiliation{%
  \institution{University of Oulu}
  \city{Oulu}
  \country{Finland}
  \postcode{90570}
}

\author{Rhema Linder}
\email{rlinder@utk.edu}
\affiliation{%
  \institution{University of Tennessee}
  \city{Knoxville}
  \country{United States}
  \postcode{37996}
}

\author{Johanna Silvennoinen}
\email{johanna.silvennoinen@jyu.fi}
\affiliation{%
  \institution{University of Jyv\"askyl\"a}
  \city{Jyv\"askyl\"a}
  \country{Finland}
  \postcode{40014}
}

\renewcommand{\shortauthors}{Oppenlaender et al.}

\begin{abstract}%
Generative AI is an emerging technology that will have a profound impact on society and individuals.
    Only a decade ago, it was thought that creative work would be among the last to be automated -- yet today, we see AI encroaching on creative domains. 
In this paper, we present the key findings of a survey study on people's perceptions of text-to-image generation.
We touch on participants' technical understanding of the emerging technology, their ideas for potential application areas, as well as concerns, risks, and dangers of text-to-image generation to society and the individual.
\chatgpt{The study found that participants were aware of the risks and dangers associated with the technology, 
but only few participants considered the technology to be a risk to themselves. Additionally, those who had tried the technology rated its future importance lower than those who had not.}
\end{abstract}%




\maketitle

\section{Introduction}%
%

\chatgpt{Recent advancements in generative artificial intelligence (GenAI) have yielded significant progress in various domains. 
This has the potential to greatly impact a wide range of industries, particularly in the creative domain.
}%
%
%
%
\chatgpt{%
Only a decade ago, the general consensus was that knowledge and creative work would be among the last to be automated \cite{The_Future_of_Employment.pdf,automation.pdf}. However, current developments in GenAI have} 
turned this prediction on its head \cite{OECD}.
Progress in generative AI has exploded in recent years and we increasingly see GenAI being applied in creative domains, such as arts and research.
One particularly popular domain is text-to-image generation,
as evident in generative systems that can synthesize images from text prompts, such as Midjourney \cite{Midjourney}, Stable Diffusion \cite{stablediffusion}, and DALL-E 2 \cite{dalle2}.
Outputs from state-of-the-art diffusion models are often indistinguishable from those generated by humans \cite{creativity,prompt-engineering}.
Some call this development ``AI's Jurassic Park moment'' \cite{Marcus.pdf} -- an adapt-or-die moment that could potentially result in massive job loss across many sectors.
However, many people are still oblivious to the generative powers of state-of-the-art systems.

In October 2022, we conducted a survey study on the 
\chatgpt{%
perception of the text-to-image generation technology among different groups of individuals, including artists as well as people with no prior experience and those who self-reported having experience with the technology.}
The survey focused on people's understanding of the emerging technology, its potential uses, and the dangers of the technology for the individual and society.
We present the key findings of the survey in this paper.



\section{Related Work}






\chatgpt{%
Related research has explored various aspects of artificial intelligence and its use in visual art.
These studies have focused on understanding the perception and attitudes towards art generated by AI \cite{AIartworkvsHumanartwork.pdf}, authorship, agency, and intention in AI-generated art \cite{prompting-ai-art,978-3-030-16667-0_3.pdf}, and the potential bias towards such art \cite{3334480.3382892.pdf}.
Additionally, \citeauthor{ncw_89.pdf} discussed the ethical implications of AI in the creative application of computer vision 
\cite{ncw_89.pdf} and
\citeauthor{PIIS2589004220307070.pdf} discussed the question of who should be credited for AI-generated works \cite{PIIS2589004220307070.pdf}.
Our paper provides a novel empirical perspective on this related work.}%
%
%
%
\section{Method}%
In autumn of 2022, we invited visitors to complete an online survey at the Researchers Night, a local annual event in which researchers present their research to the public.
The questionnaire consisted of 26 questions, including three open-ended items.
Participation was incentivized with a raffle for three Amazon vouchers, each worth 30 EUR.
We qualitatively analyzed the responses to the three open-ended survey items using in vivo coding \cite{Charmaz}. The first author read and then iteratively coded all responses.
Multiple codes were assigned, if needed, and iteratively improved by 
visualizing the codes in histogram charts.
Due to the manageable amount of data and straight-forward answers, the coding did not require multiple raters and an analysis of inter-rater reliability \cite{McDonald_Reliability_CSCW19.pdf}.%
%
%
\section{Participants}%
\chatgpt{%
35 participants (P1--P35, aged 19 to 50, $M=33.7$ years, $SD=9.3$ years) completed the online survey. 
Participants had diverse educational backgrounds, the most common being computer science, literature, and information systems. Fourteen participants held a Bachelor's degree, 10 held a Master of Science, 4 held a Master of Arts, 3 held a doctoral degree, and one held no academic degree. Twenty-four participants were students.
A third (34.3\%) of the participants had used text-to-image generation before. The most popular system used was DALL-E Mini/Craiyon (used by 7 participants), followed by DALL-E 2 (5 participants), Dream/Wombo (3 participants), and Stable Diffusion (2 participants). Participants estimated they had written an average of 20 prompts ($Max=80$, $SD=22$).
Participants were, therefore, inexperienced with the emerging technology. Participants who had tried text-to-image generation were younger than those who had not tried the technology ($p<0.05$). Ten participants considered themselves artists and had created paintings, digital art, drawings, writing, and other forms of art.%
}%
%
%
%
\section{Key Findings}%
%
%
\subsection{Understanding and Application Areas of Text-to-image Generation}%
\begin{figure}[!htb]
\centering
\begin{subfigure}[b]{0.49\textwidth}
  \centering
  \includegraphics[width=.8\textwidth]{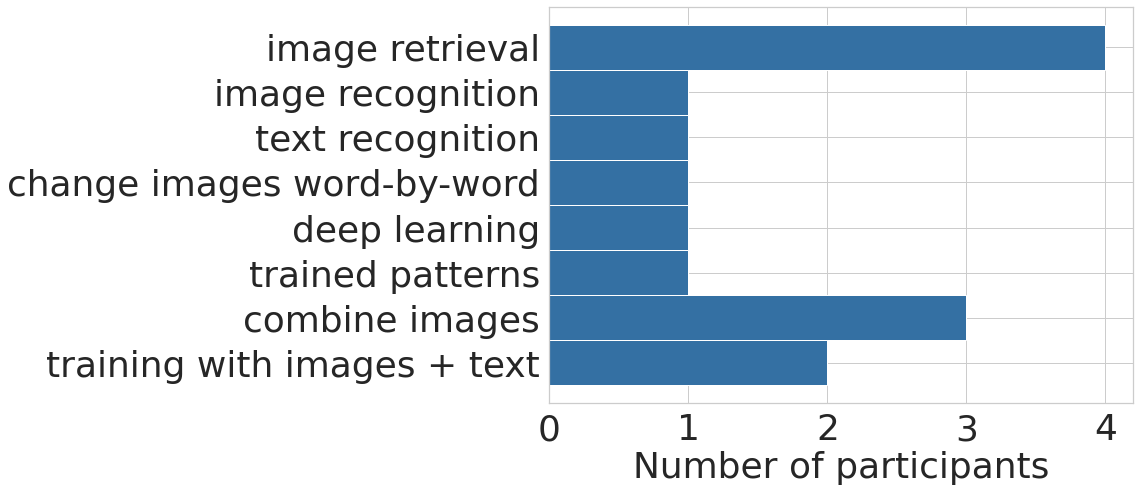}
  \caption{Participants' understanding of text-to-image generation.}
  \label{fig:working}
\end{subfigure}
\hfill
\begin{subfigure}[b]{0.49\textwidth}
  \centering
  \includegraphics[width=.8\textwidth]{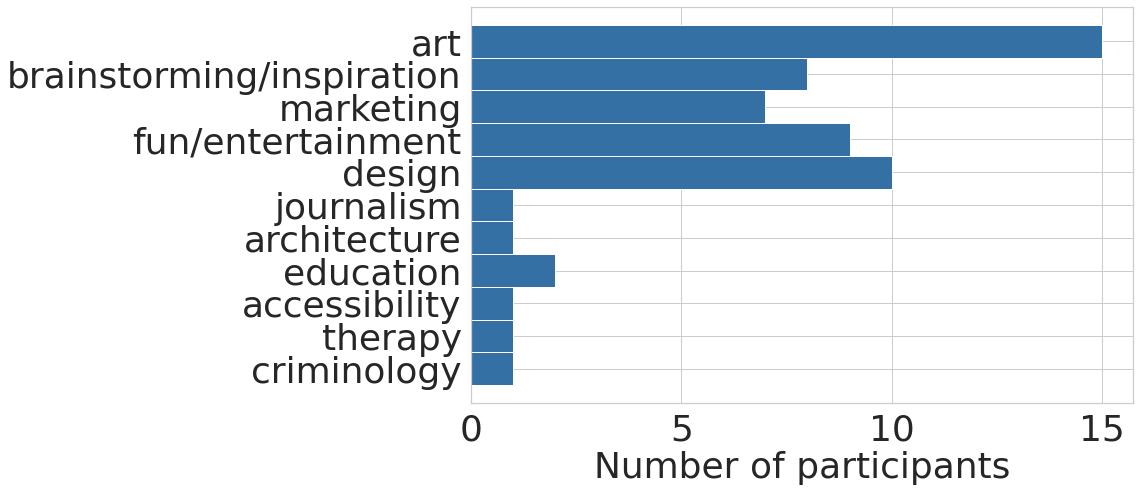}
  \caption{Participants' thoughts on potential application areas.}
 \label{fig:application}
\end{subfigure}%
\\
\begin{subfigure}[b]{0.49\textwidth}%
     \centering
     \includegraphics[width=.82\textwidth]{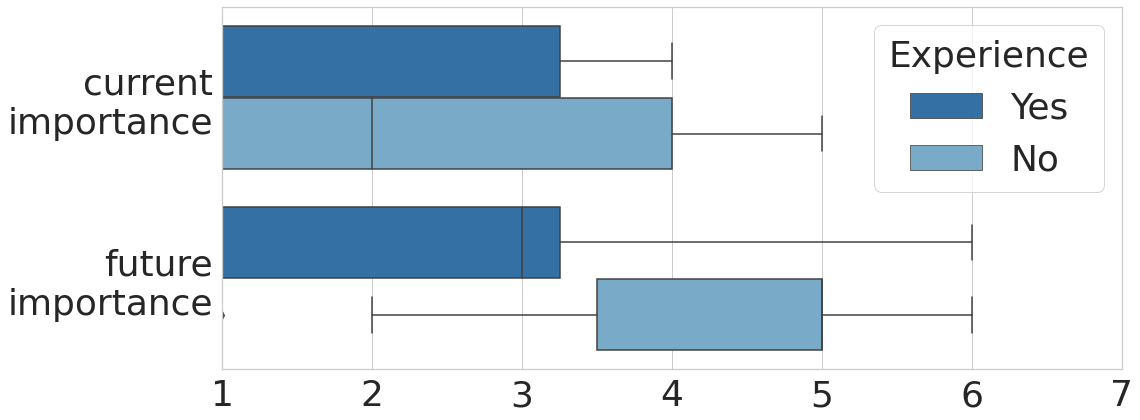}
     \\
     \includegraphics[width=.77\textwidth]{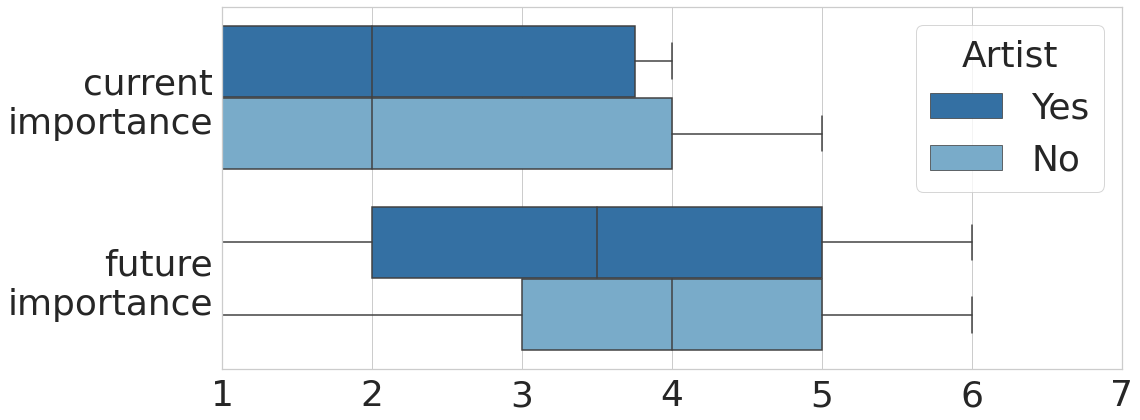}
     \caption{Current and future professional importance of image generation for participants who did and did not try text-to-image generation before (top) and self-identified artists and non-artists (bottom), from 1 -- Not At All Important to 7 -- Extremely Important.}
     \label{fig:importance}
\end{subfigure}
\hfill
\begin{subfigure}[b]{0.49\textwidth}
     \centering
    \includegraphics[width=.85\textwidth]{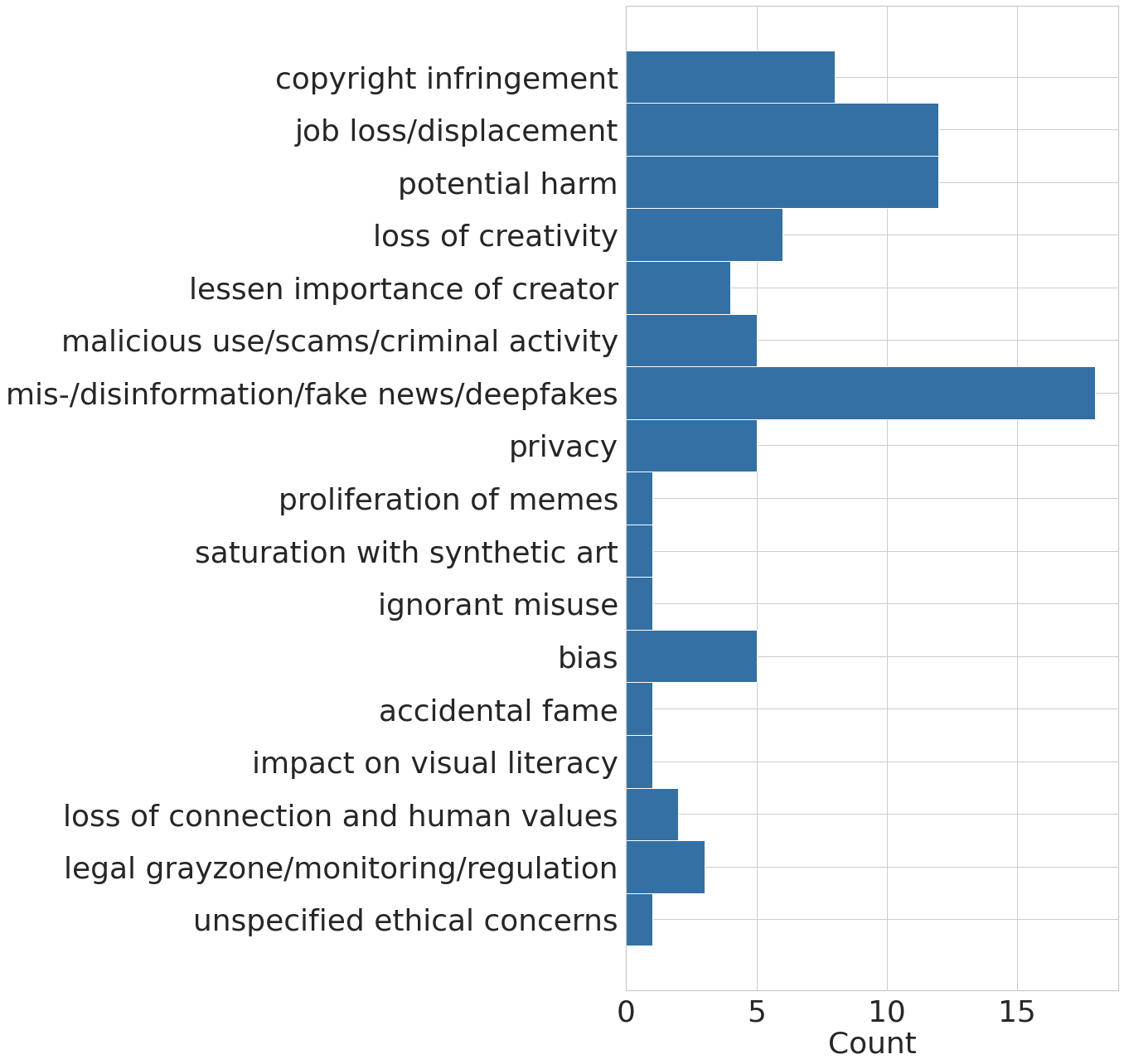}%
    \caption{Participants' concerns about text-to-image generation.}%
     \label{fig:dangers}%
\end{subfigure}
\caption{%
\chatgpt{%
Key results of the study on participants' perceptions of text-to-image generation, including their understanding of the technology, potential applications, perceived importance in their profession, and concerns.%
}%
}%
\Description{Key results of the study on participants' perceptions of text-to-image generation, including their understanding of the technology, potential applications, perceived importance in their profession, and concerns.}%
\label{fig:importancedangers}%
\end{figure}

%
Most participants did not have a strong understanding of how text-to-image generation works.
In their description of the technology, all but four participants did not distinguish between training and inference (i.e., image generation).
Participants most often related the technology to image retrieval, followed by combining or mixing existing images (see \autoref{fig:working}).
%
Creative areas dominated when it comes to participants' thoughts about potential application areas for text-to-image generation (see \autoref{fig:application}).
Participants thought the technology was 
suited for creating artworks, illustrations, and other visual media.
These could be applied in brainstorming or product development, but also marketing and design.
The entertainment industry was also seen as an application area, for instance to make animations and games.
The technology would also make a fun pass-time, according to participants.
Less common, but still interesting, application areas included therapy, education, journalism, and criminology \textit{``to reconstruct crime scenes''} (P33).


    

    

\subsection{Criticisms and Concerns about Text-to-image Generation}%

\subsubsection{Professional importance of text-to-image generation}
Most participants responded that text-to-image generation is currently not important for their profession, but will play an increasingly important role in the future (see \autoref{fig:importance}).
Interestingly, those who had tried image generation before found text-to-image generation not as important for their professional future, as opposed to those who had not tried it before. This difference was significant ($p<0.05$) and not found among self-declared artists.

\subsubsection{Concerns about the emerging technology}
The majority of participants did not think that text-to-image poses a personal danger to themselves. But participants voiced many concerns about the effect of this emerging technology on society (see \autoref{fig:dangers}).
%
%
The use of AI-generated imagery for opinion manipulation, fake news, and ``deep fakes'' was leading cause for concern.
Many participants warned that synthetic images could be spread naively (misinformation) or for malicious purposes (disinformation).
%
%
Another concern was unemployment due to increases in productivity.
   Participants mentioned that generative AI is cheaper and faster, and this could lead companies to not commission works from humans. 
Artists and designers were said to be particularly affected.
Related to the potential loss of jobs, many participants noted that text-to-image generation operates in a legal gray zone with copyright infringement being one major concern.
    P15, for instance, mentioned that some \textit{``people have directly used the artists name in the prompt to get an image to resemble the artists work as much as possible [without consent].''}
As potential long-term effect, some participants mentioned there could be a loss of appreciation for artists and their work. 
The synthetic images could \textit{``lessen the importance of the creator and the creative act''} (P4).
Artists, \textit{``who already suffer from poor income and low appreciation''} (P22), would be particularly vulnerable.

Some participants mentioned the potential effects of text-to-image generation on individuals and culture.
For the individual, the effects could include harm, such as depression and 
other illnesses related to mental health.
The AI could be used to produce offensive, abusive, and inappropriate images that are \textit{``not sensitive to people beliefs''} (P3).
The harm could be accidental, such as the negative effects of accidental fame and leakage of private information, but also intentionally abusive, such as cyberbulllying. 
As for society, several participants thought there could be a \textit{``decline in human creativity''} (P2, P23, P27).
    The technology \textit{``could curtain artist imagination, when an AI can create art better than humans''} (P3).
However, the AI was thought to be \textit{``ultimately limited in its aesthetics''} (P20). This low diversity in synthetic imagery could contribute to \textit{``narrow the viewpoint of the world''} if \textit{``a lot of images start to look the same''} (P15). 
Synthetic images could lead to a \textit{``biased and one-sided visual culture''} (P23).
As P20 noted, generative AI \textit{``has a great danger of enforcing certain values''} by showing ``mainly white European bodies with certain aesthetics.''
(P15).
The low diversity in synthetic imagery was seen to have a potentially negative impact on visual literacy if \textit{``school book visualizations are [made] in the future with AI''} (P20).
As P26 put it, text-to-image generation is \textit{``a movement away from the things that make us human, e.g, human emotions being reflected in human-made art. The knock-on effects of unemployment, depression, cause by this lack of connection with human values and needs to create and be creative''} (P26).%
\begin{figure}[!htb]%
\centering%
\begin{subfigure}[b]{0.56\textwidth}%
     \centering%
     \includegraphics[width=.7\textwidth]{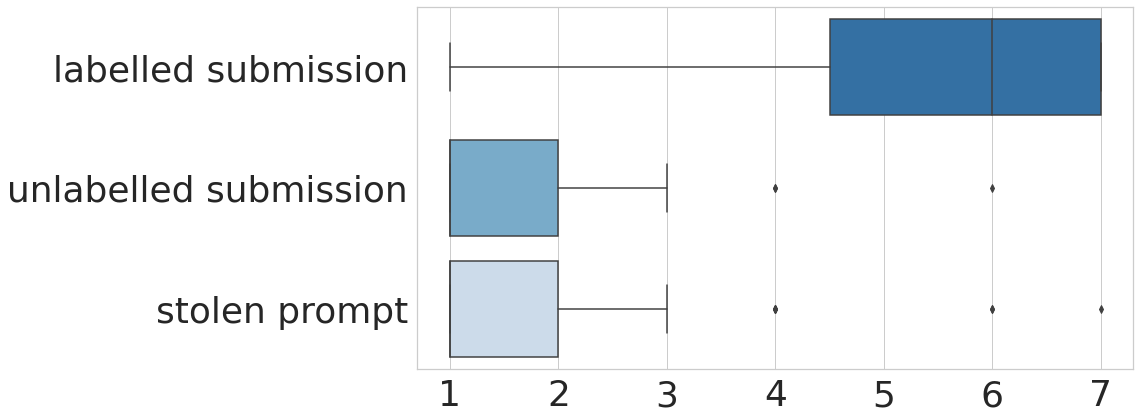}
     \caption{Rating of the ethicality of submitting a text-to-image artwork to an art fair on a Likert scale from 1 -- Not Ethical At All to 7 -- Highly Ethical.}
     \label{fig:disclosure}
\end{subfigure}
\hfill
\begin{subfigure}[b]{0.40\textwidth}
     \centering
     \includegraphics[width=.57\textwidth]{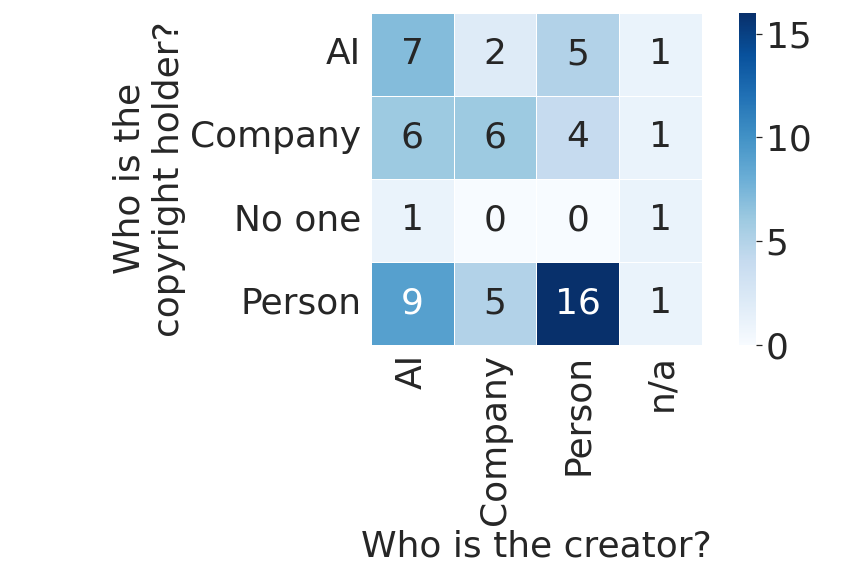}
     \caption{Participants responses to the multiple-choice questions of ``who is the copyright holder and the creator?''}
     \label{fig:copyright_creator}
\end{subfigure}
\caption{Participants thoughts on the ethicality 
(left) as well as copyright and creatorship (right).}
\Description{A woman and a girl in white dresses sit in an open car.}%
\end{figure}%
\subsubsection{Ethicality of disclosing AI generation.}%
About half of the participants ($n=19$; 54.3\%) were of the opinion that it should be disclosed when something was created with AI.
Ten participants had no strong opinion about this, and six participants (16.7\%) thought that AI-generated 
images do not need to be labeled as such.
However, when presented with the scenario of a person submitting a digital artwork to an art contest, participants thought that it was unethical to submit without disclosing that the image was created with AI (see \autoref{fig:disclosure}).
Not labeling a submission as created by AI was seen as equally unethical as submitting an artwork created from somebody else's prompt.

\section{Discussion and Conclusion}%

\chatgpt{%
Our survey study aimed to understand people's perceptions of text-to-image generation technology.
While participants did not see immediate harm for themselves, they had varied opinions on the implications of the technology for society.
}%
It seems that when it comes to pinpointing the risks and dangers of text-to-image generation as an emerging technology, it was easier for participants to enumerate the potential problems of other people
    as compared to self-reflectively analyzing the impact of 
    the technology on their own life.
\chatgpt{%
Interestingly, participants who had tried the technology rated its future importance lower compared to those who had not tried it.
    This observation conforms with the general hype cycle of technology in which expectations of an emerging technology undergo a trough of disillusionment before the technology's potential is realized \cite{HYPECYCLE}.
Our findings suggest that while there is some awareness of the emerging technology, more education and awareness is needed to help people understand the capabilities and potential implications of text-to-image generation and generative AI.
}%

\bibliographystyle{ACM-Reference-Format}
\bibliography{paper}


\end{document}